\documentclass[120t,preprint]{aastex}

\def\tighttable{\def\baselinestretch{1.1}}
\newcommand {\ltsim}{\raisebox{-.5 ex}{$\;\stackrel{<}{\sim}\;$}}
\newcommand {\gtsim}{\raisebox{-.5 ex}{$\;\stackrel{>}{\sim}\;$}}
\def\arcsec{''}
\def\asec{\ifmmode ''\!. \else $''\!.$\fi}
\def\arcsecpoint{\ifmmode ''\!. \else $''\!.$\fi}
\def\micron{\ifmmode \mu{\rm m} \else $\mu$m\fi}

\def\Msun{M$_{\odot}$}

\def\HST{{\it HST}}
\def\sst{{\it Spitzer}} 
\def\I{{F814W}}
\def\Y{{F105W}}

\def\J{{F125W}}
\def\JH{{F140W}}
\def\H{{F160W}}
\def\cl{MACS1149}

\def\obj{MACS1149-JD}
\newcommand {\lya}{Ly$\alpha$}

\shorttitle{LBGs in MACS J1149.5+2223}
\shortauthors{Zheng et~al.}
\slugcomment{To appear in the {\it Astrophysical Journal}}
\usepackage{multirow}
\articleid{AAS02490}{} 

\begin{document}

\title{Young Galaxy Candidates in the Hubble Frontier Fields \\IV. MACS J1149.5+2223}

\author{
Wei Zheng\altaffilmark{1},
Adi Zitrin\altaffilmark{2,3,4}, 
Leopoldo Infante\altaffilmark{5},
Nicolas Laporte\altaffilmark{6},
Xingxing Huang\altaffilmark{7},
John Moustakas\altaffilmark{8},
Holland C. Ford\altaffilmark{1},
Xinwen Shu\altaffilmark{7,9},
Junxian Wang\altaffilmark{7},
Jose M. Diego\altaffilmark{10},
Franz E. Bauer\altaffilmark{5,11,12},
Paulina Troncoso Iribarren\altaffilmark{5},
Tom Broadhurst\altaffilmark{13,14},
and
Alberto Molino\altaffilmark{15,16}
}
\altaffiltext{1}{Department of Physics and Astronomy, Johns Hopkins University, Baltimore, MD 21218}
\altaffiltext{2}{Cahill Center for Astronomy and Astrophysics, California Institute of Technology, MS 249-17, Pasadena, CA 91125}
\altaffiltext{3}{Physics Department, Ben-Gurion University of the Negev, P.O. Box 653, Be'er-Sheva 84105, Israel}
\altaffiltext{4}{Hubble Fellow}
\altaffiltext{5}{Instituto de Astrof\'{\i}sica and Centro de Astroingenier\'ia, Facultad de F\'isica, 
  Pontificia Universidad Cat\'olica de Chile, Santiago 22, Chile} 
\altaffiltext{6}{Department of Physics and Astronomy, University College London, London NW1 2PS, United Kingdom}
\altaffiltext{7}{Department of Astronomy, University of Science and Technology of China, Hefei, Anhui 230026, China}
\altaffiltext{8}{Department of Physics and Astronomy, Siena College, Loudonville, NY 12211}
\altaffiltext{9}{Department of Physics, Anhui Normal University, Wuhu, Anhui, 241000, China}
\altaffiltext{10}{IFCA, Instituto de F\'isica de Cantabria, UC-CSIC, s/n. E-39005 Santander, Spain} 
\altaffiltext{11}{Millennium Institute of Astrophysics, Santiago 22, Chile} 
\altaffiltext{12}{Space Science Institute, Boulder, CO 80301}
\altaffiltext{13}{Department of Theoretical Physics, University of Basque Country UPV/EHU, Bilbao, Spain}
\altaffiltext{14}{IKERBASQUE, Basque Foundation for Science, Bilbao, Spain}
\altaffiltext{15}{Instituto de Astronom\'ia, Geof\'isica e Ci\^encias Atmosf\'ericas, Universidade de S\~ao Paulo, Cidade Universit\'aria, 05508-090, S\~ao Paulo, Brazil}
\altaffiltext{16}{Instituto de Astrof\'isica de Andaluc\'ia - CSIC, Glorieta de la Astronom\'ia, s/n. E-18008, Granada, Spain}

\begin{abstract}
We search for high-redshift dropout galaxies behind the Hubble Frontier Fields (HFF) galaxy cluster MACS J1149.5+2223,
a powerful cosmic lens that has revealed a number of unique objects in its field. Using the deep 
images from the {\it Hubble} and {\it Spitzer} space telescopes, we find 11 galaxies at $z >7$
in the MACS J1149.5+2223 cluster field, and 11 in its parallel field. 
The high-redshift nature of the bright $z\simeq 9.6$ galaxy 
MACS1149-JD, previously reported by Zheng et al., is further supported by non-detection in the extremely 
deep optical images from the HFF campaign. With the new photometry,
the best photometric redshift solution for MACS1149-JD reduces slightly to $z=9.44 \pm 0.12$. 
The young galaxy has an estimated stellar mass of $(7 \pm 2) \times 10^{8}$ \Msun, and was formed at 
$z = 13.2_{-1.6}^{+1.9}$ 
when the universe was $\approx 300$ Myr old. Data available for the first four HFF clusters 
have already enabled us to find faint galaxies to an intrinsic magnitude of $M_{UV}\simeq -15.5$, 
approximately a factor of 10 deeper than the parallel fields.  
\end{abstract}


\section{INTRODUCTION}

The galaxy cluster MACS J1149.5+2223 \citep[$z=0.54$,][\cl\ hereafter]{ebeling} is one of the most interesting 
and best studied cosmic lenses. Its strong lensing power is demonstrated by many sets of multiply-imaged galaxies, 
including the largest known lensed images of a face-on spiral galaxy at $z=1.491$, whose highly magnified images 
display little distortion \citep{zitrin,smith}.
The first multiply-imaged supernova was discovered in the lensed images of this spiral, 
forming an Einstein cross \citep{kelly}. The supernova reappeared a year later in another 
counter image of the spiral as predicted by lensing models \citep{kelly2,rodney,treu}.

Because of its high magnification power, \cl\ was selected for the CLASH program 
\citep[The Cluster Lensing And Supernova survey with Hubble,][]{postman}, which 
obtained \HST\ images of 25 galaxy clusters in 16 broad bands between $0.2-1.7$ $\mu$m to a 
depth of AB magnitude $\simeq 27$. A relatively bright, young galaxy, M1149-JD at 
$z\simeq 9.6$ and magnitude $\simeq 26$,
was discovered approximately one arc minute north of the cluster center \citep[][Z12 hereafter]{zheng}. 
Lensing models based on early data \citep{zitrin, zitrin11} suggested that the source was magnified 
by approximately 
a factor of 15, and that no counter images were expected (although of note, the models lack 
sufficient constraints in this region).

\cl\ is also one of the six targets of the {\it Hubble Frontier Fields} program
\citep[HFF;][]{lotz}, whose chief goal is reaching the deepest universe with the aid
of gravitational lensing. This is particularly 
important for galaxies at $z \gtsim 7$ as they are believed to be the main source for the ionization of the 
intergalactic medium (IGM) during this period \citep{cdm,bouwens08,bouwens10,bouwens12,atek15,robertson15}, 
but most of them are too faint to be detected in random deep fields. 
In Cycle 21, Abell 2744 and MACS J0416.1$-$2403 (MACS0416) were observed, and in Cycle 22 MACS J0717.5+3745 
(MACS0717) and \cl\ were observed. In Cycle 23, Abell~S1063 and Abell~370 were observed.  Thanks to the additional, 
intensive monitoring observations of the multiply-imaged supernova (PIs: Kelly, Rodney), the 
\HST/WFC3-IR data of \cl\ are the deepest among all clusters.

The HFF clusters have therefore been the subject of many studies searching for high-redshift Lyman-break 
galaxies \citep[LBG;][]{atek14,atek15,ishigaki,mcleod,kawamata,livermore}. An important 
result is the faint-end luminosity function (LF) of galaxies down to an
intrinsic absolute magnitude of $M_{UV} \simeq -15.5$, as estimated around the rest-frame wavelength of 1500 \AA.
This is considerably deeper than the results from the Hubble Deep Field  \citep{bouwens15}, manifesting the 
uniqueness of the HFF program. 

Our group has been carrying out a systematic study of high-redshift LBGs in the HFF clusters,
reported  in a series of papers to date (\citealt{a2744,zitrin14} on Abell 2744;  
\citealt{m0416} on MACS0416; and \citealt{m0717} on MACS0717).  Here, we extend our search to \cl\  and report 11 
candidates at $z >7$ in the field of \cl\ and 11
in its parallel field.

\section{DATA}\label{sec:data}

The HFF observations of \cl\ (GO/DD 13504, PI: Lotz) were carried out
between 2013 November~2 and 2015 May~19. Additional archival images were obtained
from three groups of resource: (1) the early images (GO 9722, PI: Ebeling; GO 10493, 
PI: Gal-Yam); (2) the CLASH multi-band images (GO 12068, PI: Postman); and (3) recent WFC3/IR images
(GO~14041, 14199, PI: Kelly; GO 13790, PI: Rodney; and GO 13459, PI: Treu). 
Table~\ref{tbl-sum} lists the exposure times and limiting magnitudes for the 
imaging data used in our analysis.
 
\begin{figure} \plotone{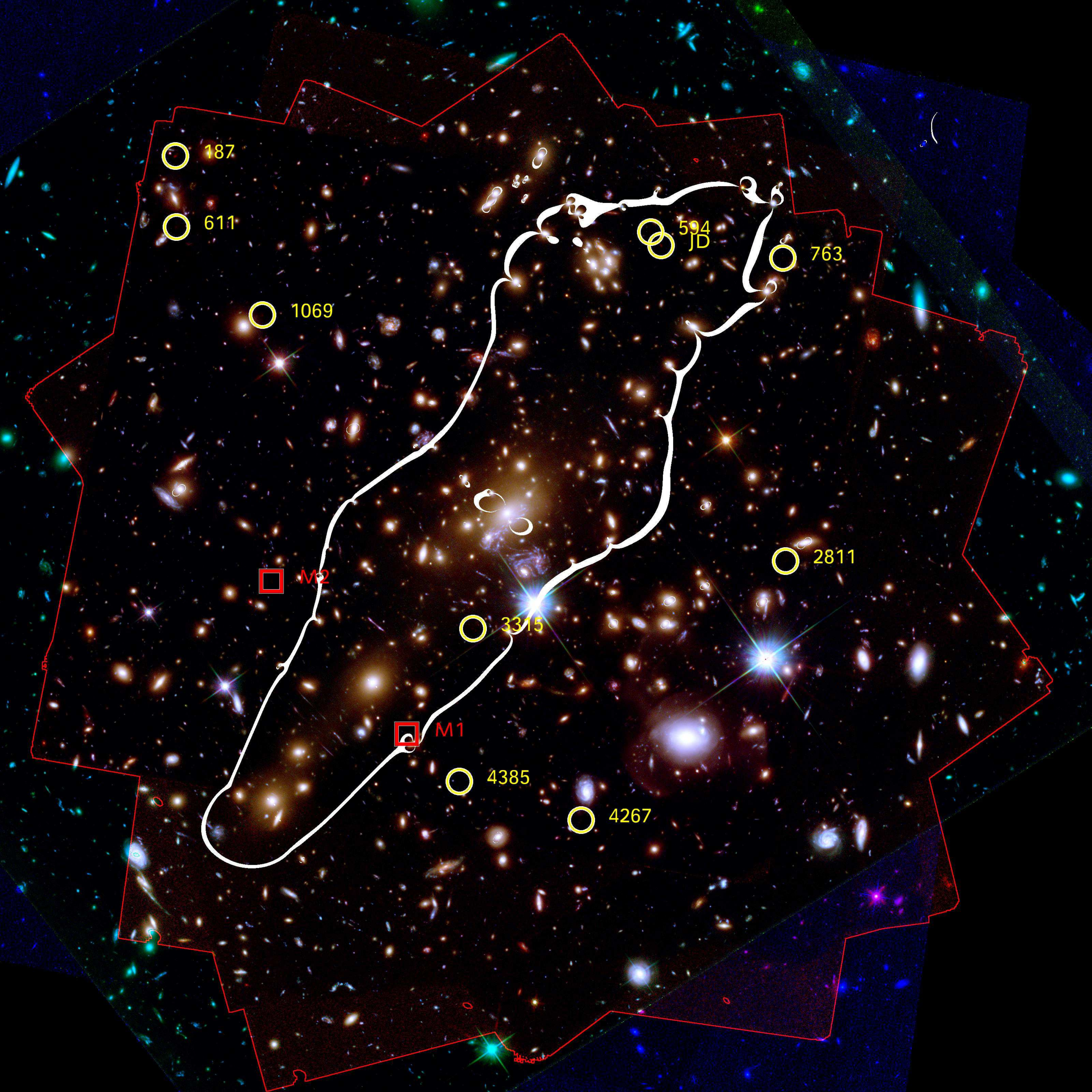}
\caption{Composite color image of \cl, made from the optical ACS images and the WFC3/IR 
images. The critical curves are from the {\sc Zitrin} {\tt LTM} model (see \S \ref{sec:lens})
for background sources at $z = 7$ and are plotted in white, marking the region with extreme 
magnification $\mu > 100$. The FOV covered by WFC3/IR is marked by a red polygon.
Yellow circles mark the LBGs at $z > 7$, and red boxes mark a possibly related $z\simeq 7.1$ 
system.                    
}\label{fig-fov}
\end{figure}

\begin{figure} \plotone{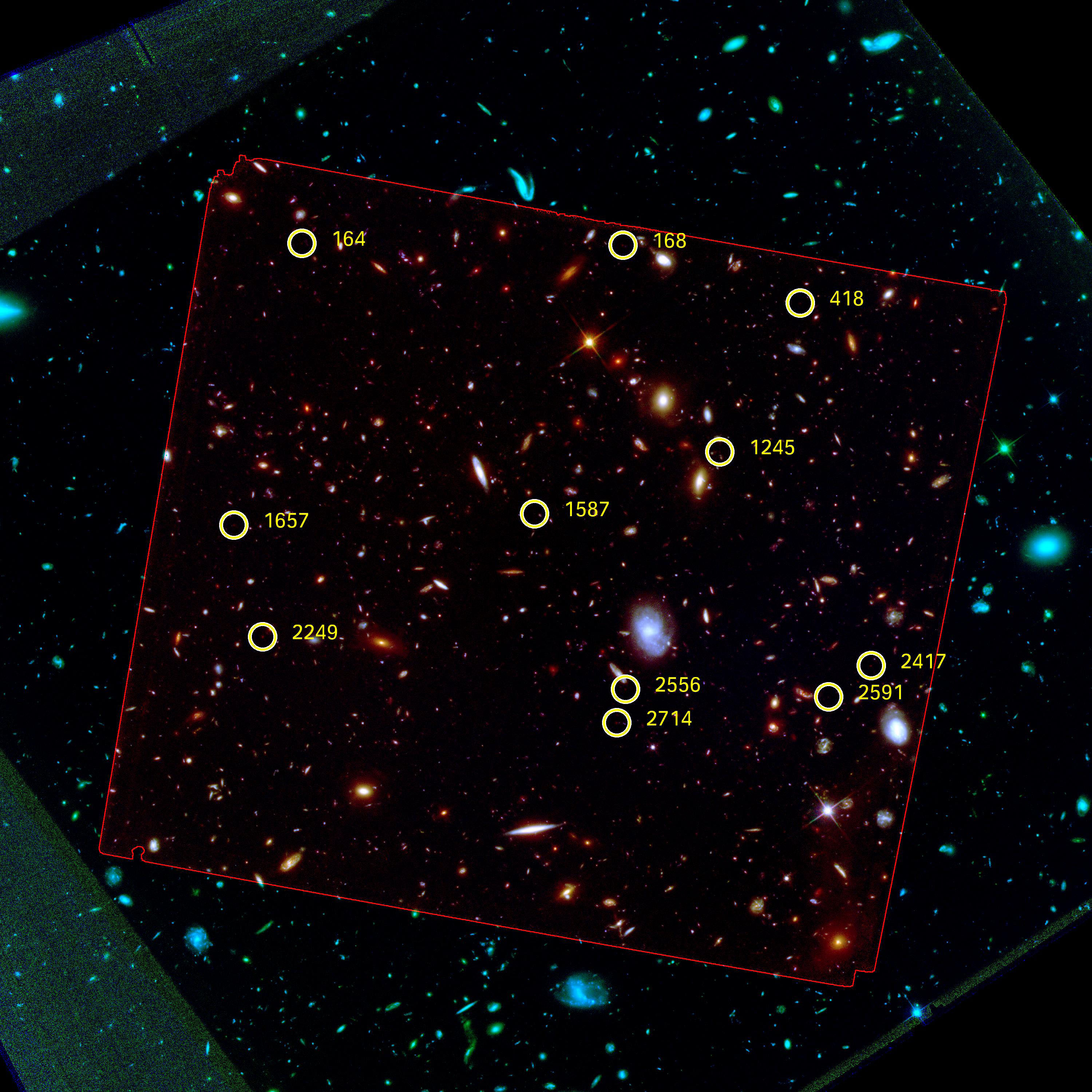}
\caption{
Composite color image of the \cl\ parallel field, made from the optical ACS images and the 
WFC3/IR F140W image. The FOV of WFC3/IR is marked by a red box, and yellow symbols 
mark the LBGs at $z > 7$.
}\label{fig-par}
\end{figure}

We processed the \HST\ data using {\tt APLUS} \citep{aplus}. 
The calibrated images from the \HST\ instrument pipelines, namely the {\it flc} images for
ACS (corrected for the detector charge transfer efficiency) and {\it flt} images for WFC3/IR, were retrieved.
The pixel scale for the final mosaic images is 0\arcsecpoint 065.  
Figure~\ref{fig-fov} displays a composite color image of the cluster field, and Figure~\ref{fig-par} 
the parallel field. The source catalogs are generated with {\tt SExtractor} \citep{bertin} in dual mode,
using the summed WFC3/IR image as a reference. As a comparison, 
we ran {\tt SExtractor} with the public HFF images \citep{hff} and checked with the measurements in Table 
\ref{tbl-cls}. The results confirm that the photometry of our candidates agrees within the stated errors.

As part of the HFF campaign, deep \sst/IRAC images of \cl\ were
obtained between 2014 Mar. and 2015 Mar. in Channels 1 and 2, using Director's
Discretionary Time (Program 90260, PI: Soifer). 
The effective exposure time in each channel, including that of the archival data (Program 60034, 
PI: Egami and 90009, PI: Brada\^c) obtained in 2010, 2011 and 2013, is $\simeq 358$~ks. The IRAC corrected 
Basic Calibrated Data (cBCD) images were processed with {\tt MOPEX} \citep{mopex} 
and sampled to a final pixel scale of $0\arcsecpoint 6$.
The estimated $1 \sigma$ limiting magnitude is $27.3$ for IRAC channel 1 (IRAC1, 3.6\micron) and $27.4$
for channel 2 (IRAC2, 4.5\micron). More details of the processing of \HST\ and IRAC data can be found in \cite{a2744}.

\section{SELECTION}\label{sec:sel}

Our selection consists of two steps: a color selection and a photometric redshift selection.
We first searched for LBGs using their distinct color around $0.1216 
(1+z)$ \micron.  At $z> 7$, these sources are optical dropouts. At $z\simeq 7 - 8$, the 
\lya\ break is at $\simeq 1$ \micron,
between the F814W and F125W bands. Our selection criteria for $z\simeq 7 - 8$, in units of magnitude, are as follows: \newline
$\I - \Y > 0.8$\newline
$\Y - \J < 0.6$\newline
$\I - \Y > 0.8 + (\Y - \J)$\newline

For $z\simeq 8 - 9$, the break is at $\simeq 1.15$ \micron, 
between the F105W and F140W bands: \newline
$\Y - \JH > 0.8$\newline
$\JH - \H < 0.6$\newline
$\Y - \JH > 0.8 + (\JH - \H)$

For candidates at $z>9$ the break is between the F125W and F160W bands:
\newline $\J - \H > 0.8$. 

We require that a candidate is not detected above $1 \sigma$ in a summed image blueward of
the selection bands defined above. For objects at $z\simeq 7$, this
requires a non-detection in a summed image of the F606W and F435W
bands, while for candidates at $z > 8$ this requires a non-detection in the stacked optical detection 
image. For candidates at $z>9$, a non-detection is also required in the F105W or even the F125W band.

In addition to the color selection criteria described above, we 
excluded candidates lying within one arcsecond of the detector edges,
in order to mitigate potentially spurious detections. Approximately 150 candidates were initially selected 
in the cluster field, and we visually inspected them. 
A large portion of these candidates were excluded as artifacts following a visual inspection, and often 
also due to an incomplete image coverage in certain bands. 
Some others are excluded as they are near stellar diffraction 
spikes, which are difficult to remove because the HFF WFC3/IR exposures were obtained at the same position
angle. Through these steps the number of candidates was reduced to approximately 30, enabling us to
carry out IRAC photometry manually with reasonable effort. The candidates with a color decrement of 
F160W - IRAC1 $> 3$ were rejected, as they are most likely extremely red objects at lower redshifts 
\citep{ero}. A similar procedure was performed in the parallel field. 

The IRAC images of our candidates suffer from crowding due to the
instrument's large point spread function (PSF, FWHM
$\simeq 1\arcsecpoint 6$), such that simple aperture photometry might
result in inaccurate fluxes due to contamination from nearby
sources. To address this issue, we used a deblending technique \citep{eyles,stark,overzier} whereby
contaminating neighbors are subtracted, by 
performing {\tt GALFIT} \citep{galfit} fits to the objects of interest and all
their close neighbors simultaneously in a fitting window of
$\sim 10\arcsec\times10\arcsec$ around the source of interest.
The IRAC PSFs were created from the isolated stars in the same image. For each source, we 
chose the PSF from nearby stars with a balance between brightness and isolation. 

Positions and radial profiles 
of neighboring sources in the region around a candidate were derived from
the \HST\ F160W-band mosaic and fixed (except for those of the bright and extended galaxies) after 
adjusting a small offset between the {\it HST} and IRAC images. 
All other parameters were allowed to vary within preset ranges. 
The initial input magnitudes were obtained by running {\tt SExtractor} on the IRAC images and partitioned by the 
flux ratios of sources in the {\HST} images. 
If one Sersic model did not yield a good fit to an extended source, an additional PSF or Sersic component was 
added in fitting.
If a source is dominantly bright, we masked out several central pixels to avoid {\tt GALFIT} being overwhelmed 
by these bright pixels. 
The typical $\chi$-square values of our fitting results are around 
10 and can be as good as 2 when nearby sources are well subtracted. The
photometric uncertainties were derived based on the fluctuations of residual images. We
performed aperture photometry in the residual images with a diameter of 2\arcsecpoint 5 and an
aperture correction of a factor of 2 and measured the flux variations as our photometric
uncertainties.

\begin{figure} \plotone{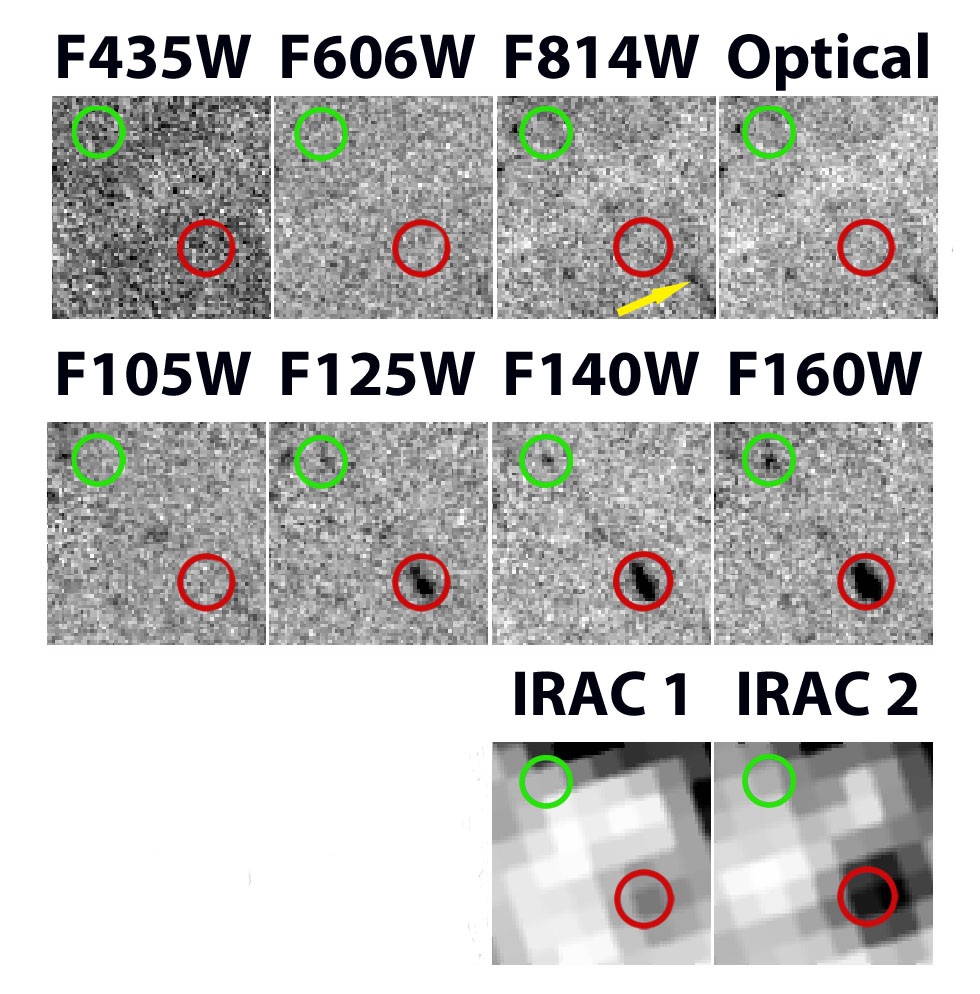}
\caption{
Cutout images of bright $z = 9.44$ LBG \obj. The object \obj\ (663) is marked by red circles, 
and the faint object 594 by green circles. 
A yellow arrow in the F814W band marks a potential contaminating source. 
The FOV is 4\arcsecpoint 6, north is up and east to the left. 
}\label{fig-jd}
\end{figure}

\begin{figure} \plotone{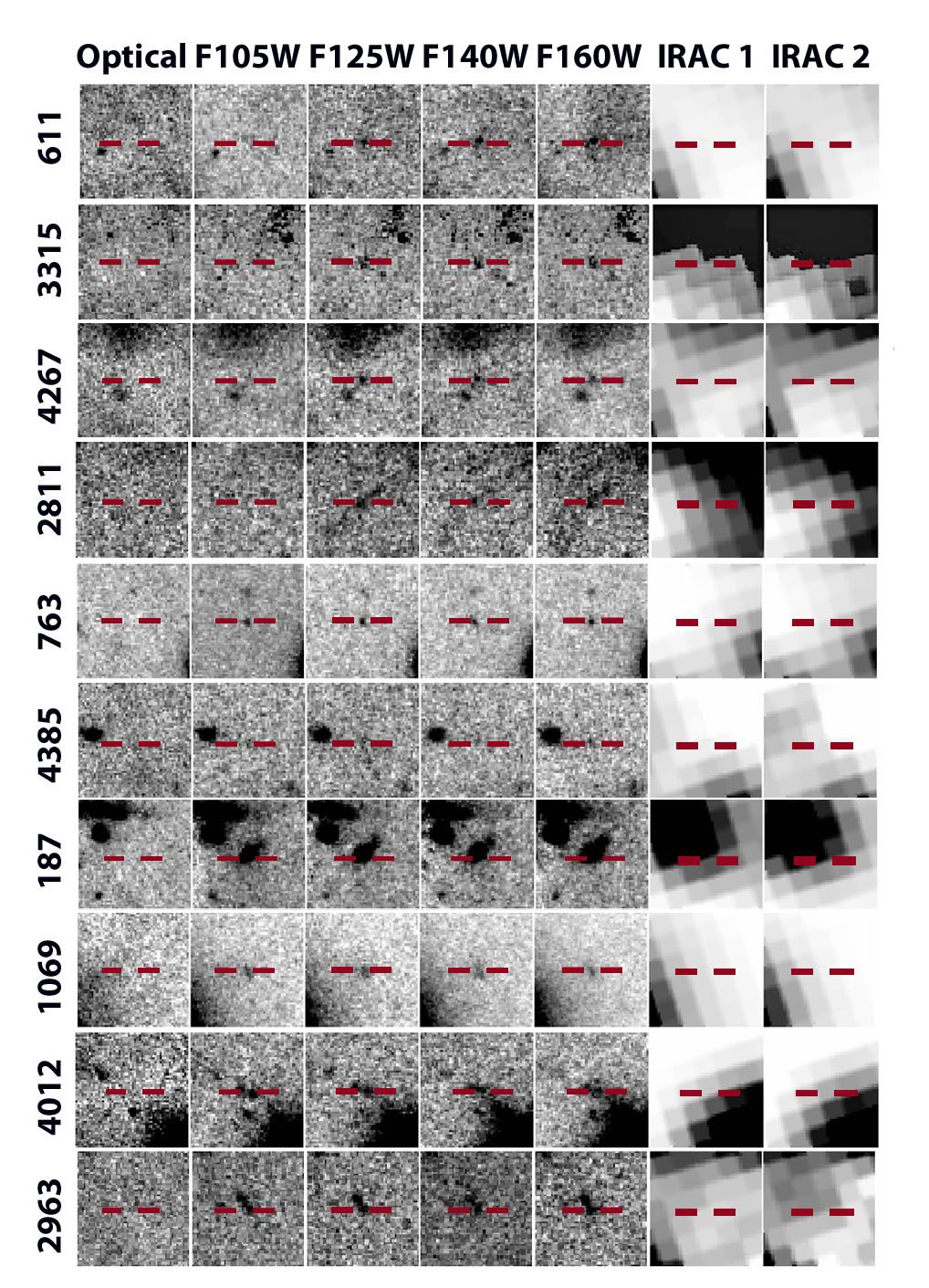}
\caption{Cutout images of other LBG candidates of $z > 7$ in \cl. 
The optical images are from the respective ACS detection images, which are
the weighted sums of ACS data in the F814W, F606W and F435W bands. 
Each candidate is at the image center, marked by pairs of red bars.
The FOV is 3\arcsecpoint 3, north is up and east to the left.
}\label{fig-stamp1}
\end{figure}

We then carried out our second-step selection, on the initial list of color-selected objects.
Using the photometry in seven \HST\ bands and two IRAC bands, we calculated photometric 
redshifts with the code {\tt BPZ} \citep[Bayesian Photometric Redshifts;][]{bpz,coe06}, 
adopting the same template library used by the CLASH collaboration \citep{jouvel}. 
We assumed flat priors on both galaxy type and redshift in the range $z=0-12$.
We required that all the candidates 
have a best-fit redshift solution of $z=7$ or higher, a $1 \sigma$ redshift range greater than 5, 
and that the probability of $z<5$ is less than 10\% of that for the high-redshift solutions.
In Figures~\ref{fig-jd}  and \ref{fig-stamp1}  we show cutout images of the $z > 7$ candidates 
in the cluster field, and in Figure~5  those of the parallel field.

Using {\tt BPZ}, we identify 25 candidates in the two fields that satisfy our color selection criteria and whose photometric redshifts (best values)
place them at $z > 7$.  In Tables~\ref{tbl-cls} and \ref{tbl-par} we list their 
coordinates, photometric redshifts, photometry, and magnifications. Four of these 25 are considered as 
the multiple images of a single galaxy at $z \simeq 7.1$ (see M1 and M2 in Table \ref{tbl-cls}). 
Therefore, we consider 22 galaxy candidates in these two fields.
Our IRAC analysis yields results for 17 sources, for which photometry
or upper limits from {\tt GALFIT} were possible. 

\section{LENS MODELS}\label{sec:lens}

As part of the HFF initiative, seven strong-lensing models and one weak-lensing model were developed for \cl\ and publicly 
released through the MAST archive. We used the strong-lensing models for \cl\ as they were made to
higher spatial resolutions. 
To estimate the systematic uncertainty in the magnification of each of our high-redshift candidates, we
calculated the median magnification factor for each candidate from the seven models (Table~\ref{tbl-cls}) and 
its corresponding range for the five middle-ranked values, namely excluding the highest and lowest magnification factors. 
This procedure is intended to mitigate potential extremes in the model predictions, and to properly reflect the underlying 
systematic uncertainties.

Figure~\ref{fig-fov} shows the composite color image of the \cl\ field, overlaid with the
critical curves from the recent {\sc Zitrin} {\tt LTM} (Light-Tracing-Mass) model and 
identification numbers for our candidates. For more details on the {\sc Zitrin} {\tt LTM} 
model see the HFF webpage and \cite{zitrin15}. Note that the critical curves in this 
best-fit model suggest that counter images for \obj\ at 
$z\sim 9.4$ should appear, although we do not secure such an identification in the 
data, and the prediction is weak given the lack of constraints in that region of 
the cluster. In other words, the exact position of the critical curves in this 
northern region is uncertain \cite[see also][K16 hereafter]{kawamata}. 
We discuss this further in \S \ref{sec:dis}. Additionally, the predicted positions of some counter images are close to 
a member galaxy near the critical curve, making predictions in that area difficult and 
sensitive to small changes in the mass of this member galaxy.

The objects in the parallel field are far away from the lensing field, but still experience
weak-lensing effects. Using the weak-lensing model of J. Merten, we derived magnification
factors for the sources in Table \ref{tbl-par} between 0.95 and 1.2. 

\section{DISCUSSION}\label{sec:dis}

\subsection{MACS1149-JD}\label{sec:jd}

The $z\sim10$ dropout MACS1149-JD, detected by Z12 using shallower CLASH \HST\ 
imaging and shallower IRAC data, remains robust. Now that considerably deeper \HST\ and 
\sst\ data are available, we revisit the properties of this object. As shown in Table 
\ref{tbl-cls}, it remains undetected in the F105W, F606W, and F435W bands, but 
weakly detected in the F814W band. We inspected the F814W image and found that the 
residual flux is not centered at the source.
There appears to be a region of extended emission of approximately 1 arcsec in size 
toward the west direction. This extended emission appears to be associated with
a source approximately one arcsecond south-west away (marked with a yellow arrow in Figure
\ref{fig-jd}). It is so weak that it can be 
seen only in the F814W and F105W bands. Even with this possible contamination, the color 
decrement between F140W and F814W is at least
3.5 mag, further reducing the possibility of
a low-redshift nature. We ran {\tt BPZ} with the new data, and found a slightly updated 
redshift of $z=9.44 \pm 0.12$,
in good agreement with our previous estimate of $z= 9.6 \pm 0.2$.
The probability of being a low-redshift object is less than $10^{-10}$.
\obj\ is therefore photometrically reconfirmed within an accurate redshift range.
The source is well detected in both the IRAC1 and 2 bands, allowing an improved estimate 
of its age and star formation rate (SFR).

The seven predicted magnification factors of \obj\ are: 9.18 for {\tt CATS} \citep{richard}; 
12.73  for {\tt Lenstool} 
\citep{johnson}; 8.48  for {\sc Zitrin}-{\tt LTM}; 11.56 for 
Zitrin {\tt LTM-Gauss} \citep{zitrin13}; 
17.0 for {\tt GLAFIC} \citep{ishigaki}; 4.8 for {\tt GRALE} \citep{grillo}; 
and 14.9 for Brada\^c \& Hoag \citep{bradac09}.  
The new models yield estimates of its magnification factor as $11.5 _{-3.1}^{+3.3}$.

\obj\ was not detected in IRAC1 in Z12 because of the shallow IRAC data that 
were available. With deeper IRAC data, \cite{bradac14} and \cite{huang} found the source's
IRAC1 magnitude as $25.7 \pm 0.5$, close to our measurement. The latest IRAC data enable an
accurate estimate of the source's intrinsic properties in the optical bands. The flux in IRAC2 is brighter
 than IRAC1 by $\approx 0.9$ mag, likely the result of a Balmer break.

In order to derive the galaxy's physical properties, we used the modeling code {\tt iSEDfit} \citep{ised} to fit 
the source's spectral energy distribution (SED). With a Monte Carlo technique, we generated $20,000$
model SEDs with a broad range of star formation histories, ages,
stellar metallicities, dust contents, and nebular emission-line
strengths. Our models in particular included nebular emission lines
whose luminosity is tied self-consistently to the number of
hydrogen-ionizing photons. For more details of {\tt iSEDfit} see Z12, \cite{a2744} and \cite{m0416}.

Figure \ref{fig-sed} displays the results for 
\obj, with a fixed {\tt BPZ} value of $z=9.44$. The derived SFR is $1.5 \pm 0.2\ 
(10/\mu)$ \Msun\ per year, with stellar mass of $6 \times 10^8 - 10^9 (10/\mu)$ \Msun, 
where $\mu$ is the magnification factor. The most interesting parameter is probably the 
galaxy age of $185 \pm 60 $ Myr, implying a formation redshift of $z\simeq 13.2 \pm 1.7$. 
The significant decrement between IRAC1 and 2 are largely explained by a Balmer decrement. The color
between IRAC1 and F160W represents the UV continuum slope, which appears to be flat in 
$F_\nu$ and shows no sign for dust extinction.
 
Object 594 in the vicinity is consistent with $z\simeq 9$. 
If physically related, they are separated by about one arcsecond (4.4 kpc in the source plane). 
While it is possible that these are counter images of the same object (see K16) due to 
lensing, it is unlikely: If the critical curve at $z \simeq 9.5$ is located between these two
 objects, they should be of roughly similar brightness and, in any case, the surface brightness must be 
the same.
Pairs of multiple images should also have been found for nearby background galaxies at lower redshifts.
The {\tt LTM} model predicts counterpart images, but none were found. Some other lensing models, 
including the previous versions of {\sc Zitrin} {\tt LTM} model, place \obj\ outside the $z=9$ critical
curve so that no multiple images are predicted. It should be noted that there is lack of 
multiple image constraints in the northern part of the critical curves and thus its extent in 
this region is poorly constrained. Given the depth of the HFF data, it becomes unlikely that 
\obj\ is multiply imaged.

\subsection{Multiple Images}\label{sec:redarc}

To help corroborate the high-redshift nature of our candidates in the cluster field, we
searched for potential counter images near the locations predicted by
the {\sc Zitrin} {\tt LTM} gravitational lensing model as well as that of \cite{diego}.
Objects M1 and M2 (see Figure \ref{fig-fov}) display similar redshifts and are at the
positions accurately predicted by the lensing models. Both images are binary, as shown 
in the lower two rows of Figure \ref{fig-stamp1}. 
Because of the high magnification, the intrinsic separation between the components is about 
0\arcsecpoint 2, or
0.8 kpc. Each of these components is unresolved even under significant magnification. 
We carried out image deconvolution of M2 using the Lucy-Richardson algorithm in 20 iterations, as 
the sources are in a region clear of contaminations.
The reference PSF image was derived from a field star. The resultant image sizes are $\ltsim 1.5$ pixel
or 0\arcsecpoint 1 in half-light radius. From a magnification factor of 7, we derived an intrinsic 
magnitude of 29.8 and $\log r_{hl} \simeq -1.4$. The result, together with that for \obj\ (Z12), is consistent with the expectation from a size-luminosity relation $r_{hl}
\propto L^{-0.5}$ \citep[Figure 9,][]{bouwens16}. 
The lensing models also predict a third image, but its position is uncertain, and no counterpart 
source is found with confidence. For faint counter images predicted near bright cluster 
galaxies (BCG), we used BCG-subtracted images to carry out our search.
Several potential candidate dropouts are seen nearby, albeit none of these pass our 
selection criteria.

Our lensing models predict that object 3315 is highly magnified and has at least two counterpart images.
One of the potential candidates is at R.A.=177.39724 and 
Decl.=22.39125. While it displays a color similar to that at $z\sim 8$, its faint magnitude of 29 does 
not yield a high {\tt BPZ} value. Another image is predicted in the northeast part of the field 
(near object 1069) but likely undetected because of the low magnification. 

\subsection{Comparison with Other Work}\label{sec:comp}

Among the 14 sources listed in Table \ref{tbl-cls}, 11 have been reported in K16. 
Object 4385 is the faintest candidate, therefore it may fall below their detection threshold.
For candidates 4267 and 2811, the photometry of the public HFF images  
confirmed independently the {\tt BPZ} values $z\sim 8.2$ for these two candidates. 

The pairs M1 and M2 were also identified in K16 as individual components at $z\simeq 6$.
Since they are undetected or in the F814W band, their redshifts should be similar and close to 7. 
Object YJ3 in K16 was not confirmed in our analysis because a weak detection in the 
F814W band places the source at redshift between 6 and 7. It is close to a red diffuse source and may 
be subject to contamination.

\section{CONCLUSION}\label{sec:con}

We searched for high-redshift dropout candidates in the HFF cluster MACS1149. We found 22 LBG candidates at $z > 7$ 
in the cluster field and the parallel field, reaching an intrinsic UV magnitude of $\approx -15.5$. Two of 
the candidates are image pairs of a single galaxy. Three of them are detected in the \sst/IRAC images. 
We also used the new, deep data to revise the fit to the previously reported $z\sim9.6$ candidate \obj\ (Z12). 
The deeper data support the high-redshift solution for \obj\ (it is not detected in the deeper optical images), and the updated 
fitting suggests a photometric redshift of $z=9.44 \pm 0.12 $, stellar mass of $(7 \pm 2) \times 10^8$~\Msun,
star-formation rate of $\simeq 1.5 $~\Msun\ per year, and a formation redshift of $z =13.2 _{-1.6}^{+1.9}$ (age of 
$\approx 300$ Myr).
Our results show once more the usefulness of using gravitational lenses for accessing the faint, early universe. 
Aided by gravitational lensing, we have found galaxies as faint as $ 0.01 L^*$, a remarkable depth for observing galaxies at the heart of 
the reionization era, in particular, in the advent of the {\it James Webb Space Telescope}.

The work presented in this paper is based on observations made with the NASA/ESA 
{\it Hubble Space Telescope}, obtained from the Data Archive at the Space Telescope Science 
Institute, which is operated by the Association of Universities for Research in Astronomy, Inc., 
under NASA contract NAS 5-26555. It is also based on
data obtained with the {\it Spitzer Space Telescope}, which is
operated by the Jet Propulsion Laboratory, California Institute of
Technology under a contract with NASA. This work utilizes
gravitational lensing models produced by the teams led by Brada\^c, 
Ishigaki, Kneib \& Natarajan, Sharon, Williams, Merten \& Zitrin, respectively.  
Support for A.Z. is provided by NASA through Hubble
Fellowship grant HST-HF-51334.01-A awarded by STScI. 
N.L. acknowledges support from a European Research Council Advanced Grant FP7/669253.
L.I., N.L., F.E.B. and P.T.I. are in part supported 
by CONICYT-Chile grants Basal-CATA PFB-06/2007, 3140542 and Conicyt-PIA-ACT 1417.
F.E.B. also thanks CONICYT-Chile grant FONDECYT Regular 1141218 and
the Ministry of Economy, Development, and Tourism's Millennium Science
Initiative through grant IC120009, awarded to The Millennium Institute
of Astrophysics, MAS. 
J.M.D. acknowledges support of the consolider projects CSD2010-00064, AYA2012-39475-C02-01 and
AYA2015-64508-P (MINECO/FEDER, UE).
A.M. acknowledges the financial support of the Brazilian funding agency FAPESP (Post-doc 
fellowship 2014/11806-9). 
X.X.H. and J.X.W. acknowledge support from the National Science Foundation of China 
(grants  11233002,11421303) and the CAS Frontier Science Key Research Program (QYZDJ-SSWSLH006).

\hskip 1.5in

\clearpage

\clearpage

\clearpage

\clearpage

\clearpage

\begin{figure} \plotone{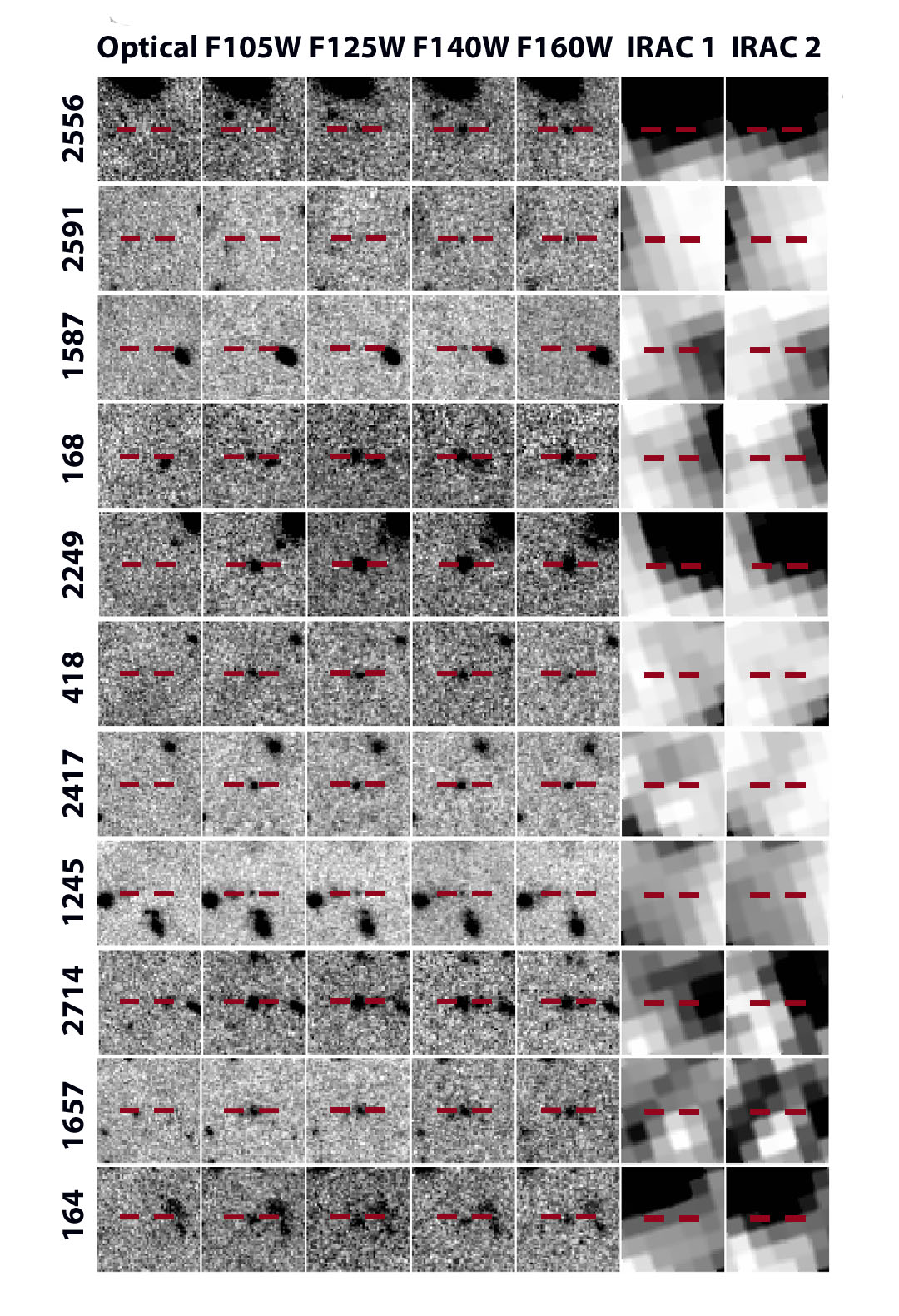}
\figcaption{Cutout images of LBG candidates 
of $z > 7$ in the \cl\ parallel field. 
The symbols are the same as Figure \ref{fig-stamp1}.
}\label{fig-stamp2}
\end{figure}
\clearpage

\begin{figure} \plotone{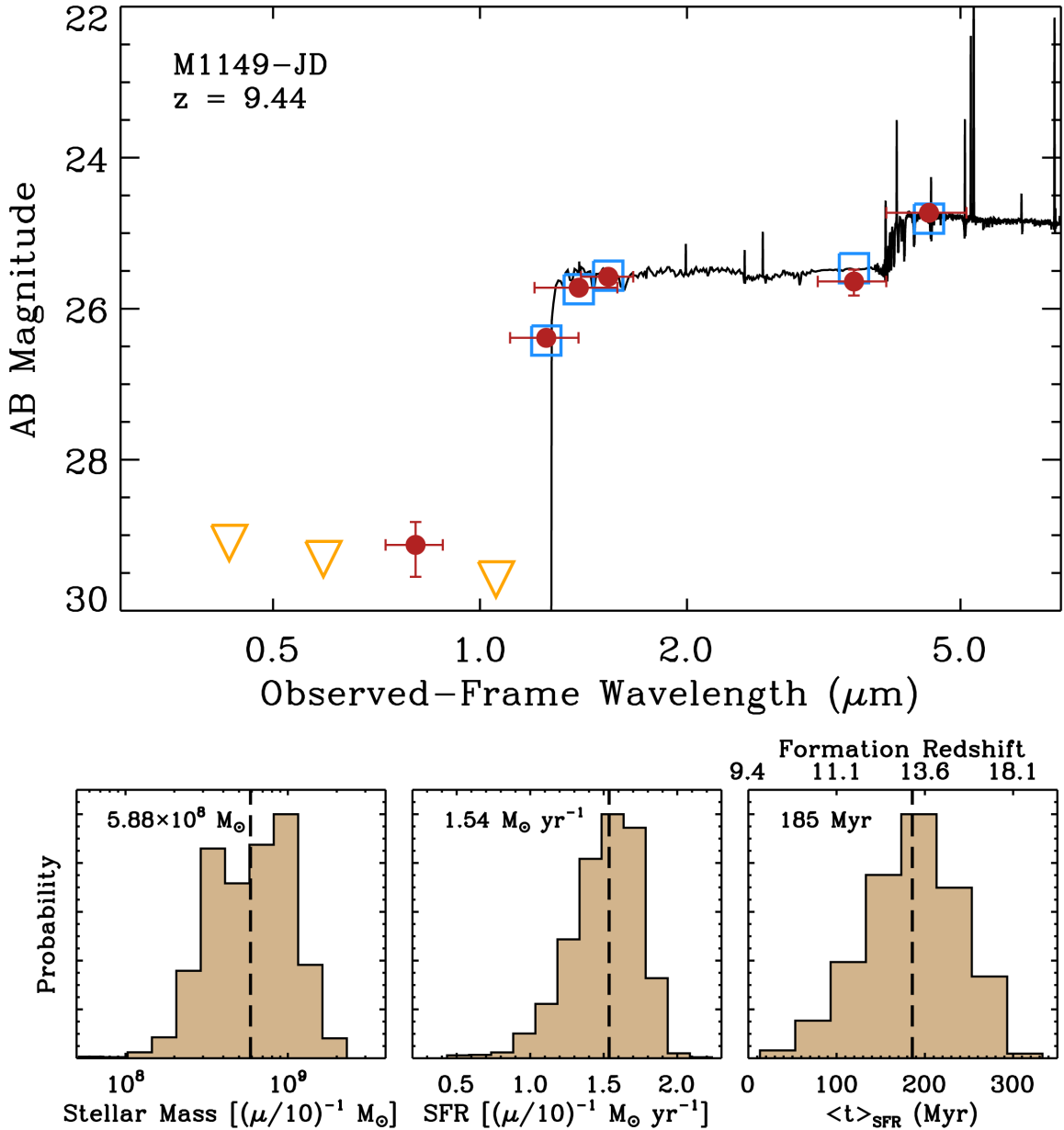}
\caption{
SED of \obj. In the top panel, the filled red points show the
observed photometry, while the open brown triangles indicate $2\sigma$
upper limits. The black spectrum shows the best-fit (maximum
likelihood) SED based on our Bayesian SED modeling using {\tt
iSEDfit}.  The blue squares show the photometry of the
best-fit model convolved with the ACS, WFC3, and IRAC filter
response curves. In the bottom panel, the probability of three fitted parameters 
are displayed.
}
\label{fig-sed}
\end{figure}
\clearpage

{
\begin{deluxetable}{cccccc}
\tablecaption{Summary of Observations\label{tbl-sum}}
\tablewidth{0pt}
\footnotesize
\tighttable
\tablehead{
\colhead{Telescope} &
\colhead{Band} &
\multicolumn{2}{c}{Date} &
\colhead{Exposure Time} &
\colhead{Limiting}
\\
\colhead{} &
\colhead{} &
\colhead{CLASH} &
\colhead{HFF} &
\colhead{(ks)} &
\colhead{Magnitude ($5\sigma$)\tablenotemark{a}}
}
\startdata \HST  & F160W&\multicolumn{2}{c}{2010 Dec.$-$2016 Feb.} 
& 107.6 & 28.9 \\ \HST  & F140W&\multicolumn{2}{c}{2011 Jan.$-$2014 Nov.}& 30.6 & 28.6 \\ 
\HST  & F125W&\multicolumn{2}{c}{2010 Dec.$-$2016 Feb.} & 66.9 & 28.9 \\ \HST  & F105W&\multicolumn{2}{c}{2011 Jan.$-$2016 Feb.} & 85.5& 29.1\\ \HST  & F814W&\multicolumn{2}{c}{2004 Apr.$-$2015 May} & 104.2 & 29.3 \\ \HST  & F606W&\multicolumn{2}{c}{2011 Jan.$-$2015 May} & 26.9 & 29.0 \\ \HST  & F435W&\multicolumn{2}{c}{2011 Feb.$-$2015 May} & 44.7 & 29.2 \\ \sst  & IRAC1&\multicolumn{2}{c}{2010 Jul.$-$2015 Mar.} & 357.7 & 25.5 \\ \sst  & IRAC2& \multicolumn{2}{c}{2010 Jul.$-$2015 Mar.}& 357.7 & 25.6 \\ 
\enddata
\tablenotetext{a}{Near the cluster center, the detection limits are lower by $\approx 0.2-0.3$ 
  mag, because of higher sky background levels from BCGs.}
\end{deluxetable}
}

\clearpage

{
\begin{deluxetable}{lcccccccccccl}
\tabletypesize{\scriptsize}
\setlength{\tabcolsep}{0.05in}
\rotate
\tablecaption{Candidates at $z > 7$ in \cl\ \tablenotemark{a}
\label{tbl-cls}}
\tablewidth{0pt}
\tighttable
\tablehead{
\colhead{Name} &
\colhead{Photometric} &
\colhead{R.A.} &
\colhead{Decl.} &
\colhead{F160W} &
\colhead{F140W} &
\colhead{F125W} &
\colhead{F105W} &
\colhead{F814W} &
\colhead{IRAC1} &
\colhead{IRAC2} &
\colhead{$\mu$\tablenotemark{b}} &
\colhead{Other}  \\
\colhead{} &
\colhead{Redshift} &
\colhead{(J2000)} &
\colhead{(J2000)} &
\colhead{} &
\colhead{} &
\colhead{} &
\colhead{} &
\colhead{} &
\colhead{} &
\colhead{} &
\colhead{} &
\colhead{Reference} 
}
\startdata
611 & $9.6 \pm 0.2$ & 177.41769 & 22.41369 & $ 26.73 \pm 0.04 $& $ 27.12 \pm 0.08 $& $ 27.75 \pm 0.15 $& $ 28.82 \pm 0.27$& $ > 31.0 $& \nodata &\nodata& $1.5 _{-0.4}^{+0.2}$ & K16 (Y2) \\ 
JD (663) & $9.44^{+0.11}_{-0.13}$ & 177.38995 & 22.41271 & $ 25.70 \pm 0.01 $& $25.88 \pm 0.02 $& $ 26.57 \pm 0.03 $& $ >30.7 $& $ 29.3 \pm 0.3 $& $25.64 \pm 0.17 $& $24.73 \pm 0.07 $ & $11.5 _{-3.1}^{+3.3}$ & Z12, K16 (YJ1) \\ 594 & $ 9.1 \pm 0.5$& 177.39055 &  22.41341 & $28.07 \pm 0.10 $& $ 28.09 \pm 0.13 $& $ 28.73 \pm 0.21 $& $> 30.7 $& $ > 31.0 $& \nodata & \nodata & $10.4 _{-2.2}^{+9.0}$ & K16 (YJ4)\\ 
3315 & $ 8.7^{+0.4}_{-0.3}$& 177.40073 & 22.39244 & $28.11 \pm 0.08 $& $ 28.13 \pm 0.10 $& $ 28.34 \pm 0.11 $& $> 30.7 $& $ > 31.0 $&\nodata & \nodata & $20 _{-14}^{+22}$ & K16 (YJ3) \\
4267 & $ 8.3 _{-0.7}^{+0.3}$& 177.39453 & 22.38231 & $ 28.29 \pm 0.09 $& $ 28.04 \pm 0.09 $& $ 28.32 \pm 0.09 $& $ 29.49 \pm 0.27 $& $> 31.0 $ &  $>26.1 $& $ >25.5 $ &$1.7 _{-0.3}^{+0.2}$ & \\ 
2811 & $ 8.0_{-0.7}^{+0.4}$& 177.38284 & 22.39600 & $ 27.75 \pm 0.09 $& $ 27.85 \pm 0.14 $& $ 27.70 \pm 0.11 $& $ 28.98 \pm 0.30 $& $> 31.0 $ & \nodata & \nodata & $2.4 _{-0.5}^{+1.4} $ \\ 
763 & $ 7.7 _{-0.6}^{+0.3}$& 177.38298 & 22.41203 & $ 28.05 \pm 0.08 $& $ 27.88 \pm 0.08 $& $ 28.03 \pm 0.09 $& $ 28.75 \pm 0.12 $& $> 31.0 $ & 
$ > 25.6 $& $ > 25.6 $ & $3.6 _{-1.3}^{+3.7}$ & K16 (i19)\\ 
4385 & $ 7.7 _{-1.3}^{+0.8}$& 177.40151 & 22.38434 & $ 28.62 \pm 0.09 $& $ 28.36 \pm 0.10 $& $ 28.72 \pm 0.10 $& $ 29.62 \pm 0.21 $& $ > 31.0 $&\nodata & \nodata & $2.6 _{-0.1}^{+1.3}$ &  \\ 
187 & $ 7.3 \pm 0.1$& 177.41776 & 22.41744 & $ 24.88 \pm 0.02 $& $ 24.87 \pm 0.02 $& $ 24.92 \pm 0.02 $& $ 25.31 \pm 0.02 $& $ 28.3 \pm 0.3 $ & \nodata & \nodata &$1.4 _{-0.3}^{+0.2}$ & K16 (i1)\\ 
1069 & $ 7.3 _{-0.7}^{+0.3}$& 177.41278 & 22.40902 & $ 27.44 \pm 0.07 $& $ 27.21 \pm 0.07 $& $ 27.45 \pm 0.08 $& $ 27.77 \pm 0.09 $& $> 31.0 $ & \nodata & \nodata & $1.8 _{-0.3}^{+0.2}$ & K16 (i11)\\ 
M1-4012\tablenotemark{c} & $ 7.2 ^{+0.3}_{-0.4}$& 177.40451 & 22.38688 & $ 27.01 \pm 0.06 $& $ 26.83 \pm 0.07 $& $ 27.08 \pm 0.07 $& $ 27.31 \pm 0.08 $& $> 31.0 $ & 
$ >25.5$\tablenotemark{d} & $ >24.9$\tablenotemark{d} & $7.0 _{-2.4}^{+39}$ & K16 (39.2)\\ M1-4023 & $ 7.2 \pm 0.3$& 177.40460 & 22.38668 & $ 26.48 \pm 0.10 $& $ 26.55 \pm 0.10 $& $ 26.88 \pm 0.13 $& $ 26.93 \pm 0.11 $& $27.7 \pm 0.2$ & $ >25.5$\tablenotemark{d}& $ >24.9$\tablenotemark{d} &$ 7.0^{+39}_{-2.4}$ & K16 (21.1)\\ M2-2952 & $ 6.9 _{-1.5}^{+0.3}$& 177.41226 & 22.39497 & $ 27.72 \pm 0.08 $& $ 27.67 \pm 0.10 $& $ 27.60 \pm 0.07 $& $ 27.72 \pm 0.07 $& $ > 31.0$ & $> 25.5 $\tablenotemark{d}& $> 26.5$\tablenotemark{d} & $6.6 _{-1.9}^{+3.1}$ & K16 (i24) \\ 
M2-2963 & $ 7.1 _{-1.1}^{+0.3}$& 177.41220 & 22.39489 & $ 27.89 \pm 0.08 $& $ 27.69 \pm 0.09 $& $ 27.78 \pm 0.08 $& $ 27.97 \pm 0.08 $& $ > 31.0$ &$> 25.5 $\tablenotemark{d}& $> 26.5 $\tablenotemark{d} &$6.9 _{-2.0}^{+3.4}$ & K16 (i26) \\ 
\enddata
\tablenotetext{a}{\HST\ magnitudes are isophotal,
scaled by an aperture correction term derived in the \H\ band. The errors and 
limiting magnitudes are $1\sigma$. Photometric redshifts have been derived using {\tt BPZ}, and the quoted
uncertainties indicate the 68\% confidence interval.}
\tablenotetext{b}{Magnification factor is the mean of seven models and the 
range of five middled-ranked models after excluding the highest
and lowest values.}
\tablenotetext{c}{{\tt BPZ} is uncertain because of contamination from a nearby object. }
\tablenotetext{d}{Upper limit for both components, which are unresolved in IRAC images.}
\end{deluxetable}
}

\clearpage

{
\begin{deluxetable}{lcccccccccc}
\tabletypesize{\scriptsize}
\setlength{\tabcolsep}{0.05in}
\rotate
\tablecaption{Candidates at $z > 7$ in the \cl\ Parallel Field
\label{tbl-par}}
\tablewidth{0pt}
\tighttable
\tablehead{
\colhead{Name} &
\colhead{Photometric} &
\colhead{R.A.} &
\colhead{Decl.} &
\colhead{F160W} &
\colhead{F140W} &
\colhead{F125W} &
\colhead{F105W} &
\colhead{F814W} &
\colhead{IRAC1} &
\colhead{IRAC2} 
\\ \colhead{} &
\colhead{Redshift} &
\colhead{(J2000)} &
\colhead{(J2000)} &
\colhead{} &
\colhead{} &
\colhead{} &
\colhead{} &
\colhead{} &
\colhead{} &
\colhead{} 
}
\startdata
2556 & $ 9.3 _{-0.4}^{+0.3}$& 177.41644 & 22.29354 & $ 27.77 \pm 0.07 $& $ 27.83 \pm 0.07 $& $ 28.52 \pm 0.15 $& $> 30.7 $& $> 31.0 $  & $> 27.1 $& $> 27.3 $ \\ 
2591& $ 9.1 _{-6.9}^{+0.6}$ & 177.40555 & 22.29315 & $ 29.10 \pm 0.13 $& $ 29.05 \pm 0.13 $& $ 29.91 \pm 0.30 $& $> 30.7 $& $> 31.0 $ & $> 27.1 $& $ > 26.9 $\\ 
1587& $ 8.8 _{-7.9}^{+0.5}$& 177.42131 & 22.30224 & $ 30.22 \pm 0.34 $& $ 29.06 \pm 0.12 $& $ 30.44 \pm 0.42 $& $ >30.7 $& $ > 31.0 $ & $>27.5  $& $ >27.3 $ \\ 
168& $ 7.9 \pm 0.2$ & 177.41658 & 22.31556 & $ 27.00 \pm 0.05 $& $ 27.14 \pm 0.06 $& $ 27.27 \pm 0.06 $& $ 28.27 \pm 0.10 $& $> 31.0 $ & $ 26.0 \pm 0.5 $\tablenotemark{a}& $ 25.4 \pm 0.2 $\tablenotemark{a}\\ 
2249& $ 7.7 \pm 0.1$ & 177.43588 & 22.29614 & $ 26.50 \pm 0.03 $& $ 26.04 \pm 0.02 $& $ 26.29 \pm 0.03 $& $ 26.96 \pm 0.03 $& $> 31.0 $ & $>27.2$& $ 24.9\pm 0.1$\tablenotemark{a}\\ 
418& $ 7.5_{-0.6}^{+0.3}$ & 177.40706 & 22.31268 & $ 27.96 \pm 0.08 $& $ 27.94 \pm 0.07 $& $ 27.92 \pm 0.08 $& $ 28.52 \pm 0.09 $& $ > 31.0$  &$ > 27.5 $&$ > 27.1 $\\ 
2417 & $ 7.4 _{-0.4}^{+0.2}$& 177.40326 & 22.29470 & $ 27.33 \pm 0.05 $& $ 27.29 \pm 0.05 $& $ 27.44 \pm 0.05 $& $ 27.85 \pm 0.06 $& $> 31.0 $ &$ 27.3 \pm 0.8 $&$ 27.4 \pm 0.6 $\\ 
1245& $ 7.2 _{-1.0}^{+0.3}$ & 177.41138 & 22.30530 & $ 28.67 \pm 0.13 $& $ 28.28 \pm 0.09 $& $ 28.53 \pm 0.11 $& $ 28.81 \pm 0.10 $& $> 31.0 $ &$ > 26.0 $&$ > 26.9 $\\ 
2714& $ 7.1 \pm 0.2$& 177.41692 & 22.29187 & $ 26.82 \pm 0.04 $& $ 26.73 \pm 0.04 $& $ 26.88 \pm 0.04 $& $ 27.09 \pm 0.04 $& $> 31.0 $ &$ > 27.2 $&$ > 27.3 $\\ 
1657& $ 7.2 \pm 0.2$ & 177.43742 & 22.30168 & $ 27.44 \pm 0.06 $& $ 27.25 \pm 0.05 $& $ 27.22 \pm 0.05 $& $ 27.52 \pm 0.04 $& $> 31.0 $ &$ > 28.0 $&$ > 27.3 $\\ 
164& $ 7.1 _{-1.0}^{+0.2}$ & 177.43379 & 22.31566 & $ 28.27 \pm 0.09 $& $ 28.30 \pm 0.09 $& $ 28.04 \pm 0.08 $& $ 28.39 \pm 0.07 $& $ 30.3 \pm 0.7 $ &$ > 27.3 $&$ > 27.3 $ \enddata
\end{deluxetable}
\tablenotetext{a}{Near a bright source. The magnitude error may be larger than the tabulated result.}
}
\clearpage

\end{document}